# Interfacial Entanglement-Induced Time-Dependent Solidification of Polymeric Fluids

Jaewon Shim (심재원)[1,†], Manhee Lee (이만희)[2,*], Wonho Jhe (제원호)[1,3*]

[1]Center for 0D Nanofluidics, Institute of Applied Physics, Department of Physics and Astronomy, Seoul National University, Seoul 08826, Korea

[2]Department of Physics, Chungbuk National University, Cheongju, Chungbuk 28644, Korea

[3]John A. Paulson School of Engineering and Applied Sciences, Harvard University, Cambridge, MA

[†]Present address: Institute of Applied Physics, Department of Physics and Astronomy, Seoul National University, Seoul 08826, Korea

[*]Corresponding author contact: whjhe@snu.ac.kr, mlee@cbnu.ac.kr

## Abstract

The structure of polymers at solid interfaces evolves over time, but the corresponding changes in their rheological properties remain poorly understood. Here, using a home-built quartz tuning fork atomic force microscope-based nano-rheometer, we directly measure the time-dependent viscoelasticity of the interfacial fluid. The bottommost layer, closest to the substrate, undergoes solidification over 10 hours, exhibiting an approximately five-fold increase in storage modulus and a two-fold increase in loss modulus. This arises from interfacial entanglement due to the strong binding of polymers to the solid surface driven by solid-wall attractive interactions. In contrast, within the second and third layers, the storage modulus remains nearly constant over time, while the loss modulus shows approximately two-fold increase. In this region, unlike the strongly bound first layer, entropic repulsion dominates, allowing the material to behave fluid-like while becoming increasingly viscous. Notably, as the first layer, where interfacial entanglement occurs, undergoes solidification, the flow boundary for interfacial fluid flow shifts upward away from the substrate, resulting in a negative slip length. This highlights the critical role of nanoscale interfacial structure and properties in governing macroscopic flow behavior.

## I. Introduction

Fluids exhibit structural organization when adjacent to solid surfaces on the nanometer scale[1–3]. This interfacial ordering is attributed to solid-fluid interfacial interactions and the physical confinement imposed by the solid boundary[4,5]. As a result, fluids develop oscillatory density distributions extending across several molecular diameters normal to the interface, with a periodicity comparable to the molecular size of the fluid[6–9]. A wide variety of fluids exhibit such interfacial layering with solid substrates, including globular liquids[10,11], polar and nonpolar molecules[12–14], and linear polymer chain fluids[15,16].

Interestingly, the interfacial structure of fluids exhibits aging behavior over time. X-ray reflectometry (XRR) studies revealed time-dependent conformational rearrangements at interfaces[17,18]. Similar temporal evolutions were observed using surface force apparatus (SFA)[19] and atomic force microscope (AFM)[20]. While these techniques provide visualization of structural changes at interfaces, the rheological properties associated with such time-dependent structural evolution remain largely unexplored.

Given that structure and properties are inherently linked[12,21,22], material properties are expected to evolve with interfacial structural changes over times. However, directly measuring these time-dependent properties remains challenging due to limitations of conventional rheometers, which are difficult to probe below a gap size of ~1 μm[23]. To access the range below 1 μm, various nanoscale rheological techniques have been developed. For instance, SFA enables rheological characterization under nanoconfinement[24], primarily capturing the average mechanical properties of thin films using a millimeter-sized probe. Thus, it is difficult to resolve local structural changes or molecular-scale relaxation dynamics[3]. Cantilever-based AFM can measure interaction forces beneath a nanoscale tip to infer interfacial fluid structure[20,25], yet extracting quantitative rheological properties from these force measurements is difficult. Moreover, both SFA and AFM typically experience jump-to-contact instability, which limits precise control of confinement distances at the sub-nanometer scale[26]. Nano dynamic mechanical analysis, a technique based on AFM, can effectively probe local rheological properties in the normal direction; however, it is not suited for characterizing shear responses[27].

Here, we employ a quartz tuning fork (QTF)-based AFM in shear mode to perform rheology of interfacial polymers on solid surfaces. The high quality factor ($Q$~8,000) and stiffness ($k_0$~40 kN/m) of the QTF sensor[28,29] allow sensitive measurements with sub-nanometer resolution along the $z$-axis in fluid environments, while avoiding jump-to-contact instability. This approach not only reveals the interfacial fluid structure but also quantitatively assesses rheological properties as a function of distance from the solid surface. Essentially, we observe that the bottommost layer undergoes solidification over time due to interfacial entanglement during physical aging. Surprisingly, this occurs even though polymer entanglement at solid interface is suppressed due to the limited conformational freedom of the polymer chains[30]. Specifically, the storage modulus ($G'$) increases by a factor of five and the loss modulus ($G''$) doubles. Meanwhile, the second and third layers, located farther from the substrate, show minimal change in $G'$, while $G''$ increases approximately twofold. After 10 hours of aging, the ratio of $G'/G''$ of the first layer increases by ~40%, indicating solidification induced by interfacial entanglement[31]. In contrast, the $G'/G''$ of the second layer decreases by 26% and that of third layers by 38%, reflecting enhanced fluidity driven by entropic repulsion[20,32]. Furthermore, the progressive solidification of the first layer leads to an upward shift of the flow boundary over 10 hours, resulting in a negative slip length.

# II. Nano-rheology of polymeric fluids

## A. Experimental setup

Figure 1(a) shows a schematic of the QTF-AFM-based rheometer. The tip is fabricated by pulling quartz capillaries and attaching it to the QTF prong using epoxy adhesive (Araldite), followed by a 12-hour curing period. The probe diameter is approximately 20 nm. The probe tip is positioned perpendicular to the substrate to shear confined fluids. A piezoelectric actuator (P-123.10, PI), installed beneath the solid substrate, provides z-axis displacement control with sub-nanometer precision (0.05 nm) under sealed AFM conditions. A function generator (33120A, Agilent) is used to acuate the QTF probe with an amplitude of 0.4–1.0 nm at a resonance frequency near 32 kHz, and a lock-in amplifier (SR830, Stanford Research Systems) is used to simultaneously measure the oscillation amplitude and phase shift. A freshly cleaved mica (Grade V-1, SPI Supplies) is prepared immediately prior to measurement. The 1 μL sample droplet of polydimethylsiloxane (PDMS)-based silicone oil (Sigma-Aldrich, 500 cSt, $M_w$ ~ 18,000 g/mol) is deposited on the mica, and measurements begin immediately after the sample deposition. All experiments are conducted at a controlled temperature of 23°C and relative humidity of 40%. The polymeric droplet was deposited onto the mica substrate and spread to a diameter of approximately 1 cm. We measured the viscoelasticity at 8 different positions by laterally displacing the probe in ~1 μm steps using motorized stages at each aging time. Despite spatial separation, the measured viscoelastic moduli and layering intervals remained consistent across all positions within the central region of the sample droplet, where temperature and concentration gradients are expected to be negligible. To ensure reliable results, we have avoided the measurements near the droplet boundary, where temperature and concentration deviation could arise due to surface curvature or spreading. Furthermore, the low volatility of PDMS-based silicone oils reduces the likelihood of concentration or temperature variations during measurement.

To sensitively probe the interfacial fluids, the tip is partially immersed in the sample fluid (Fig. 1(b)), where we obtain the QTF signals as a force-zero reference. Then, as the tip approaches the substrate, the QTF signals change, from which we obtain rheological properties of interfacial fluids as a function of tip-substrate distance. Compared to the resonance in air, the resonance with partially immersed tip in sample shows a frequency shift of 0.06 Hz and the quality factor (Q-factor) decreases by approximately 1,000 due to the damping interaction between the tip and the sample fluid. This measurement scheme has two key advantages: First, a high Q-factor ($Q{\sim}8{,}000$) of the tip in liquid enables sensitive measurement of mechanical responses. Second, the high stiffness of the QTF prong ($\sim 10^4$ N/m)[33] ensures atomic-resolution height control while avoiding mechanical instabilities such as jump-to-contact[28]. Although Brownian fluctuations are often non-negligible at molecular length scales, thermal fluctuations in the motion of QTF prongs are minimal due to the high-stiffness of the prongs. The estimated root-mean-square of Brownian displacement is approximately 0.28 pm, which is 3,500 times smaller than the tip's oscillation amplitude (~1.0 nm). Therefore, Brownian fluctuations of the probe is neglected, allowing accurate measurement of external viscoelastic interactions using the QTF probe.

A schematic of the interfacial fluid structure on the mica surface is shown in Fig. 1(c). The polymer conformation at the solid interface differs from that in the bulk[5,34] due to the attractive interaction between the confining mica wall and the silicone oil sample[35,36]. While bulk polymers consist of randomly moving melts[37], polymers near the solid interface exhibit layered structures[5,25]. This phenomenon manifests as an oscillatory density distribution, with layering intervals typically order of the molecular diameter. For polymeric fluids like PDMS, siloxane-based chains can become partially immobilized near the solid surface *via* hydrogen bonding with surface hydroxyl groups[38]. Interestingly, ordered structures of periodic surface patterns were reported for metallic fluids at solid surfaces[39]. In that case, the ordered structure was formed due to Marangoni convection under an electrostatic oscillatory field. Unlike this macroscopic ordering, our system demonstrates nanoscale ordering of polymers at the attractive solid surface.

Figure 1(d) presents the amplitude and phase signals of the QTF during approach toward the substrate. These approach curves resolve the layered structure of the interfacial polymers with sub-nanometer *z*-axis resolution. The raw data of amplitude and phase, without background subtraction, reveals three distinct peaks, indicating discrete interfacial layers. The periodicity of 0.8 nm corresponds closely to the size of a PDMS monomer. The peak positions are identified as follows: $d_0$ is 0.46 nm, approximately half the size of a PDMS monomer; $d_1$ is 0.81 nm and $d_2$ is 1.0 nm, where they closely match the full monomer size[20] (See supplementary information S1). The tip-substrate contact point is defined as the position where the dissipation energy reaches its maximum, occurring when the dynamically oscillating tuning fork rapidly dissipates its driving energy upon contact with the solid substrate. We designate this point of maximum dissipation as the zero-contact point with a precision of 0.1 nm[40].

## B. Viscoelasticity analysis

Viscoelasticity is obtained using the amplitude-modulation mode of the QTF-based rheometer. With controlled tip-substrate distance at sub-nanometer scale, we obtain the *in situ* shear response of confined fluid through the QTF sensor's amplitude and phase signals, acquired *via* the lock-in detection scheme. We model the QTF-based rheometry system as a simple harmonic oscillator, with the equation of motion given by:

$$m\ddot{x} + b\dot{x} + kx = F_0 e^{i\omega t} + F_{int}, \tag{1}$$

where $k = 49{,}399\ \mathrm{N/m}$ is the stiffness of QTF obtained from the thermal spectrum, $b$ is the damping coefficient of the system, $m$ the effective mass, $\omega$ the driving angular frequency, $F_0 e^{i\omega t}$ the driving force, and $F_{int} = -k_{\mathrm{int}}x - b_{\mathrm{int}}\dot{x}$ the interaction force with elasticity $k_{\mathrm{int}}$ and damping coefficient $b_{\mathrm{int}}\omega$. The interaction coefficients, $k_{\mathrm{int}}$ and $b_{\mathrm{int}}\omega$, are obtained by inserting the tip oscillation $x(t) = A\sin(\omega t + \phi)$ into Eq. (1)[28],

$$\frac{k_{\mathrm{int}}}{k} = \frac{A_0}{QA}\sin\phi + \left(\frac{\omega}{\omega_0}\right)^2 - 1, \tag{2}$$

$$\frac{b_{\mathrm{int}}}{b} = \frac{\omega_0 A_0}{\omega A}\cos\phi - 1, \tag{3}$$

where $A$ is the amplitude, $\phi$ is the phase, $A_0$ the amplitude at resonance, $Q$ the quality factor, $\omega_0$ the natural angular frequency. The shear elasticity $k_{\mathrm{int}}$ and damping coefficient $b_{\mathrm{int}}\omega$ characterize the viscoelastic properties of confined fluids between the tip and the substrate. In our study, we measure the silicone oil with a molecular weight of 18,000 g/mol (500 cSt) at a thickness of tens of nanometers on the mica substrate. The viscous penetration depth of the oil sample is, $\delta = \sqrt{2\eta/\omega\rho} \approx 70\mu\mathrm{m}$, where $\eta$ is the viscosity and $\rho$ is the density. Since the tip-substrate distance, tens of nanometers, is much smaller than $\delta \approx 70\mu\mathrm{m}$, the planar Couette flow assumption holds, and thus the strain is given by $\gamma = A/z$. The measured stress is then $-F_{int}/\sigma_{\mathrm{eff}} = (k_{int}A + ib_{int}\omega A)/\sigma_{\mathrm{eff}}$, where $\sigma_{\mathrm{eff}}$ is the effective probing area of the tip. From this stress-strain relation, the complex shear modulus $G^*$, defined as the ratio of stress to strain, is obtained, $G^* = G' + iG''$,

$$G' = \left(\frac{k_{\mathrm{int}}\,A}{\sigma_{\mathrm{eff}}\,\pi r^2}\right)/\frac{A}{z} = k_{\mathrm{int}}\,z/\sigma_{\mathrm{eff}}\pi r^2, \tag{4}$$

$$G'' = \left(\frac{b_{\mathrm{int}}\omega\,A}{\sigma_{\mathrm{eff}}\,\pi r^2}\right)/\frac{A}{z} = b_{\mathrm{int}}\omega\,z/\sigma_{\mathrm{eff}}\pi r^2. \tag{5}$$

Here, the storage modulus $G'$ quantifies the elastic resistance of interfacial polymeric fluids to shear deformation, while the loss modulus $G''$ represents energy dissipation under the same strain conditions[34,41]. While the mica substrate is atomically flat, the probe tip is curved. Therefore, the effective area $\sigma_{\mathrm{eff}}$ is introduced to account for the curvature of the tip[34]. Note that all viscoelastic measurements are performed at a fixed frequency of 32 kHz, which corresponds to the lateral resonance mode of our quartz tuning fork sensor[42]. This frequency was selected to maximize measurement sensitivity.

## C. Force analysis based on Maxwell viscoelastic theory

We employ the Maxwell viscoelastic model to analyze the lateral force response of confined polymeric fluids under nanoscale oscillatory shear. This model is selected to capture both elastic (storage modulus, $G'$) and damping (loss modulus, $G''$) contributions to the interfacial dynamics, which are not adequately captured by purely Newtonian descriptions under confinement[43]. Prior studies have demonstrated that even simple liquids, such as water, exhibit viscoelastic behavior when confined at the nanoscale[44,45], where the Maxwell model has been successfully applied to describe such responses.

The Maxell model, consisting of a spring and dashpot in series, provides a minimal yet sufficient framework for describing stress relaxation behavior through the characteristic time $\tau$. In the case of confined Laminar flow (detailed derivations in Supplementary Information S2), the elastic $k_{int}$ and damping coefficients $b_{int}\omega$ are given as

$$k_{int} + ib_{int}\omega = \kappa\frac{1-i\tau\omega}{1+(\tau\omega)^2}\left(f_{\tau\omega} + if_{\tau\omega}^{-1}\right)\cot\{(f_{\tau\omega} + if_{\tau\omega}^{-1})z/\delta\}, \tag{6}$$

where $\kappa = \sigma\omega^{3/2}\sqrt{\eta\rho/2}$ , $f_{\tau\omega} = \sqrt{\tau\omega + \sqrt{(\tau\omega)^2 + 1}}$ , and $\delta = \sqrt{2\eta/\omega\rho}$ with the interfacial viscosity $\eta$, the effective contact area of the tip $\sigma$, and the angular frequency $\omega$. The dimensionless parameter $f_{\tau\omega}$ characterizes the elastic response of the fluids. Notably, when $\tau\omega = 0$, the model reduces to the Newtonian case[46].

For nanoconfined conditions where $(z/\delta)f_{\tau\omega} \ll 1$, $\cot\{(f_{\tau\omega} + if_{\tau\omega}^{-1})z/\delta\} \cong \frac{1}{(f_{\tau\omega}+if_{\tau\omega}^{-1})}\frac{\delta}{z}$ and thus the corresponding viscoelastic response becomes,

$$k_{int} + ib_{int}\omega = \kappa\frac{\tau\omega+i}{1+(\tau\omega)^2}\frac{\delta}{z}. \tag{7}$$

Equation (7) shows that the shear interaction with Maxwell fluid can be decomposed into elastic and damping components. The elastic force reflects the storage modulus $G'$, while the damping force corresponds to the loss modulus $G''$.

Since our measurements are performed at a fixed frequency of 32 kHz, the resonance frequency of the QTF sensor, the ratio $G'/G''$ indicates the material's relaxation behavior, such as, solid-like when $G'/G''$ > 1 and fluid-like when $G'/G''$ < 1. By analyzing the lateral damping force at this frequency, we can characterize the distance-dependent nanoscale flow profile in proximity to the solid wall.

Equation (7) implies that the shear interaction scales with $1/z$, suggesting that the hydrodynamic flow vanishes at $z = 0$. However, in our nanofluidic system, the actual $z$-position at which the flow vanishes may vary depending on interfacial properties. To account for this, the shear damping force is generally expressed as

$$F_{\text{shear}} = \frac{\eta\,\sigma\,v_{\text{shear}}}{z+b}, \tag{8}$$

where $b$ is the slip length and $v_{\text{shear}} = \omega A$ is the shearing speed with $A$ as oscillation amplitude. By fitting experimental data with Eq. (7), we extract the slip length $b$, a key parameter that quantifies deviations from the traditional no-slip boundary condition at solid interfaces.

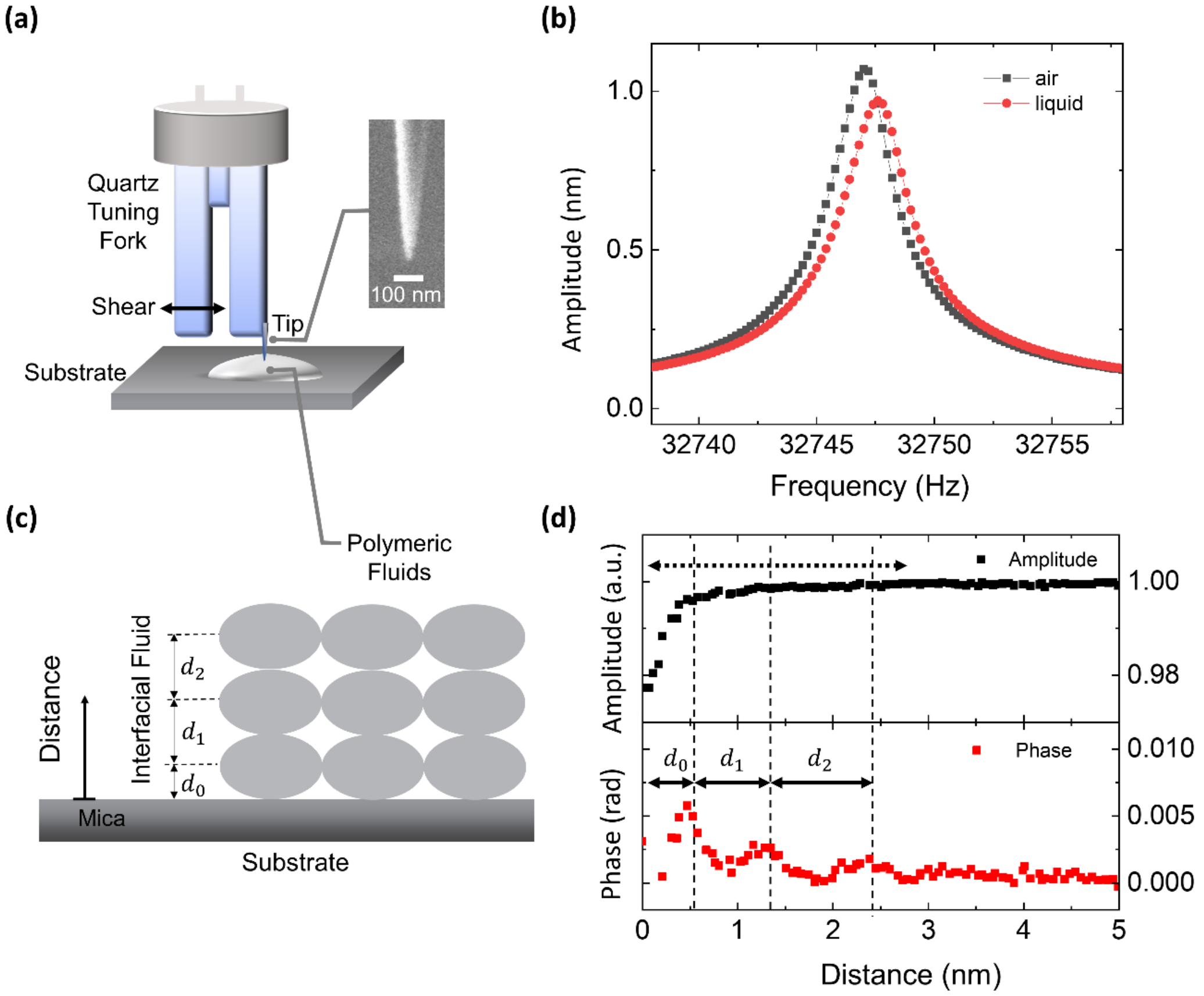

**Fig. 1.** (a) Experimental setup of QTF-AFM-based nanorheometer. The QTF sensor measures mechanical interactions by detecting amplitude and phase signals. A probe tip is partially immersed in the sample fluid. A drop of sample, about $1\mu L$ volume, is placed on the mica surface. The tip diameter at the apex is 20 nm. The scale bar represents 100 nm. (b) In-air resonance curve is measured above the fluid surface, with a resonance frequency of 32,744 Hz and $Q$-factor of 8,700. The in-liquid resonance curve is obtained at the tip-substrate distance of approximately 100 nm, with resonance frequency of 32,744.6 Hz and $Q$-factor of 7,677. (c) Schematic illustration of PDMS-based silicone oil molecules at the interface, showing an ordered layered structure distinct from the bulk arrangement. The interlayer distances, $d_1$ and $d_2$, correspond to the size of the PDMS monomer, while $d_0$ corresponds to approximately half that size. (d) Typical approach curves of the QTF signal as a function of tip-substrate distance. The black and red dots represent amplitude and phase signals, where interfacial structural signals are observed in the phase signal.

# III. Interfacial aging of polymeric fluids

Once the silicone oil droplet is deposited, a layered structure of polymers forms on the mica surface. This induces oscillatory shear elasticity $k_{int}$ and a monotonically increasing shear damping $b_{int}\omega$. Using Eqs. (4) and (5), these parameters are converted into the complex shear moduli $G'$ and $G''$, as shown in Figs. 2(a) and (b). We define polymers at $z$ =0.46 nm as Layer-1 (the first layer), Layer-2 (the second) at $z = 1.28$ nm, and Layer-3 (the third) at $z = 2.28$ nm, as indicated by dashed lines. During the 10-hour measurement period, the oscillatory behavior of $G'$ as a function of distance are observed within 5 nm from the surface (Fig. 2(a)). We define the region containing the three layers as the interfacial region. Over time, both $G'$ and $G''$ in the Layer-1, Layer-2, and Layer-3 regions continue to increase (Fig. 2(b)).

Figure 2(c) shows the storage modulus at local maxima (dashed lines) at different times. The storage modulus of Layer-2 is approximately 5 times greater than the bulk plateau storage modulus at 32 kHz ($G'_{bulk} = 0.032$ MPa)[34,47], whereas that of Layer-3 is about 3 times greater. Despite undergoing 10 hours of physical aging, both layers retain their structural shape. In contrast, Layer-1 exhibits a significant increase in $G'$, reaching a value 16 times greater than the bulk modulus after 10 hours. Additionally, the loss modulus $G''$ at each layer increases over time (Fig 2(d)). The loss modulus of Layer-3 within 1 hour is comparable to the bulk plateau loss modulus at 32 kHz ($G''_{bulk} = 0.057$ MPa)[34,47].

The initially measured complex modulus of Layer-1, Layer-2, and Layer-3 are higher than that of the bulk (Figs. 2(c,d)), which is attributed to the formation of layered PDMS polymer structures at the interface (1-hour). In our previous study[34], we observed an increase in the storage modulus of PDMS polymers near solid surfaces, using various macroscopic probe tips with diameter ranging from 10 μm to 100 μm. Notably, a 20 nm tip exhibited oscillatory responses corresponding to the size of individual polymer monomers[34], which is consistent with the initial observations in the present study (1-hour). Furthermore, as the probe approaches the substrate, viscous dissipation becomes dominat[48,49], leading to an increase in the loss modulus. The increased loss modulus implies enhanced viscosity in nanoscale proximity to the solid wall[50]. Therefore, we generally suggest that the layered structure of polymers at solid interfaces enhances viscoelasticity beyond bulk values. Over time, the layered polymers experience the adsorption-induced entanglement, leading to solidification. While this behavior may resemble that of cross-linked PDMS, such cross-linked materials typically exhibit moduli, approximately 1–3 MPa[51,52], far exceeding the values measured in our study. The cross-linked polymers are interconnected through covalent bonds[53], unlike the physically entangled polymers studied here. At short tip-sample distances below ~ 0.5 nm (approximately half the monomer size), the standard deviation of repeated measurements increases significantly. This variability can be attributed to several factors including tip geometry, enhanced molecular-scale repulsion, and elevated energy dissipation under strong confinement[48]. In addition, as the interfacial polymer layer undergoes time-dependent entanglement, the resulting nanostructure becomes increasingly anisotropic[54], leading to direction-dependent tip-surface interactions that further contribute to measurement variations.

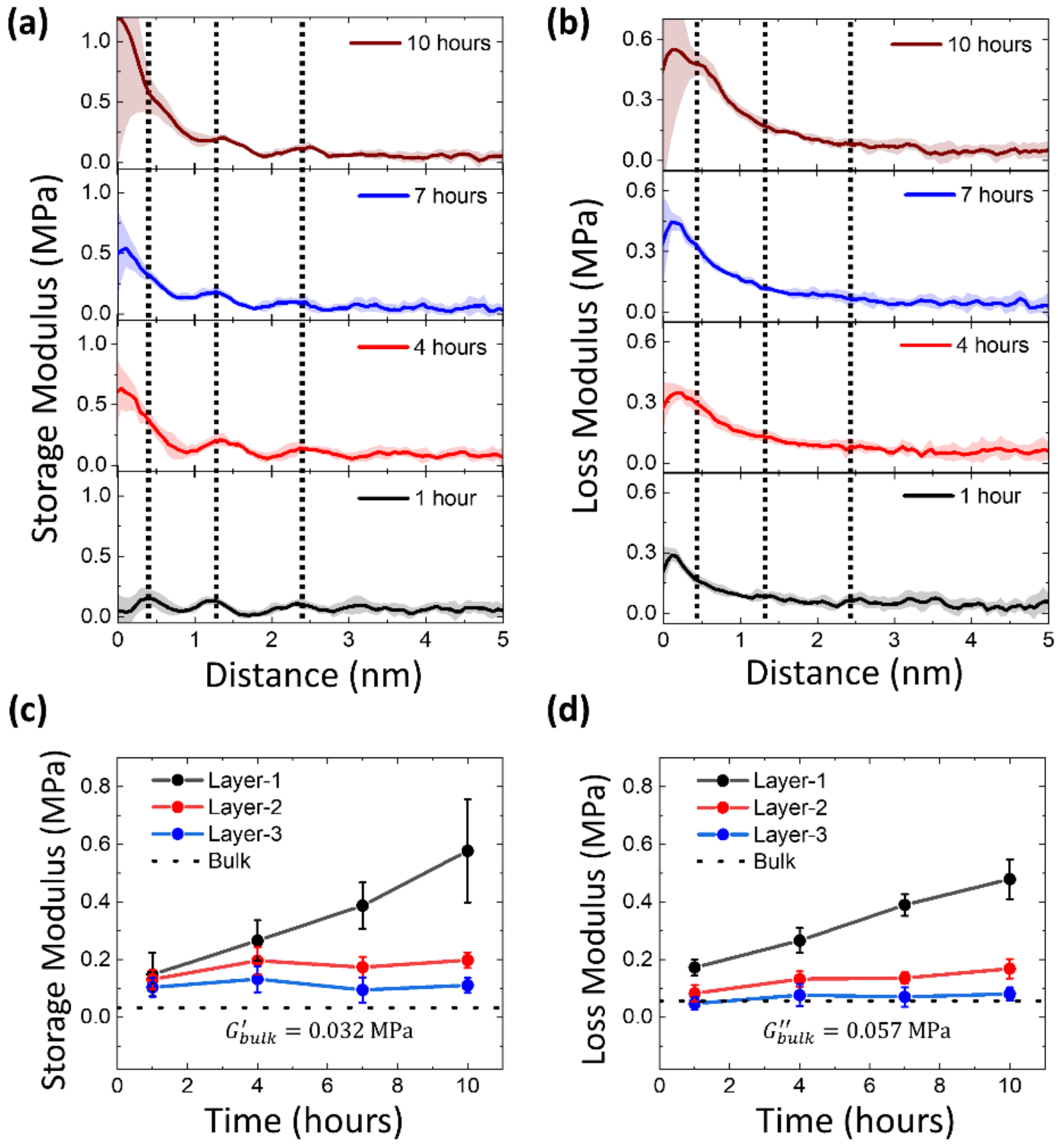

Fig. 2. (a,b) Storage modulus $G'$ and loss modulus $G''$ as a function of tip-substrate distance. Shaded areas represent the standard deviation of 8 repeated measurements. The black, red, blue, and olive curves correspond to the storage and loss moduli at 1, 4, 7, and 10 hours, respectively. (c) Storage modulus $G'$ at the peak positions of each interfacial layer at different times. Over time, the storage modulus of Layer-1 increases fivefold, while those of Layer-2 and Layer-3 remain relatively constant. This suggests that the interfacial fluid near 1 nm from the substrate becomes increasingly rigid. After 10 hours, the modulus of Layer-1 reaches approximately 18 times the bulk value ($G' = 0.032$ MPa at 32 kHz)[34]. (d) Loss moduli of all layers increase with time, indicating progressive enhancement of viscosity in the interfacial fluids. Notably, the loss modulus of Layer-1 doubles after 10 hours of aging, whereas that of Layer-3 remains comparable to the bulk value ($G'' = 0.057$ MPa at 32 kHz)[34].

Figure 3(a) illustrates the time-dependent behavior of interfacial layered polymers, highlighting the development of interfacial entanglement within Layer-1 over time. Initially, PDMS polymers form a layered structure on the mica substrate. Due to attractive interactions with the mica, polymers continuously adsorb onto the surface and interconnect with one another, resulting in interfacial entanglement. This physical aging process increases the storage modulus[31,55], leading to the solidification of Layer-1, corresponding to the fluid within approximately 1 nm from the mica surface.

The rheological state of the interfacial polymers (Fig. 3(a)) is quantified using the dimensionless mechanical relaxation time, $\tau\omega$ (Fig. 3(b)), defined as the ratio of storage to loss modulus, $G'/G''$, based on the viscoelastic Maxwell model[44,56]. Our nanorheological measurements are performed at the resonance frequency of QTF, 32 kHz, to sensitively probe the viscoelastic properties of the fluid. In the Maxwell model, the relaxation time is strictly defined at the crossover point where the storage modulus equals the loss modules, i.e., $G' = G''$. However, when measurements are performed at a fixed frequency, the ratio $G'/G''$ can serve as a proxy for estimating the relaxation behavior of the fluid. This approach is commonly employed in nanometric tip-based rheometry[7,44,57]. The ratio $G'/G''$ of each layer is normalized to its initial value at 1 hour to show relative changes over time. Error bars represent the standard deviation of 8 measurements taken at each layer peak position ±0.05 nm (see S3 of the supplementary material for details).

The ratio of $G'/G''$ values in Layer-1 behaves differently compared to Layer-2 and Layer-3. Layer-1 becomes increasingly solid-like, with an approximately 40% increase, indicating solidification driven by interfacial entanglement[55]. The solid-like response in Layer-1, located within ~1 nm from the substrate, arises from a combination of chain entanglements and attractive interactions with solid wall. In particular, hydrogen bonding between siloxane segments in PDMS and hydroxyl groups on solid surface can induce partial immobilization of polymer segments[38], promoting physical adsorption and reducing conformational freedom. With aging, early-arriving chains tend to adsorb in flat conformations, while later-arriving chains interact with an already crowded interface and adopt looped or tailed configurations. These surface-anchored chains dynamically entangle with adjacent chains, leading to the formation of interfacial entanglement with enhanced storage modulus and increased relaxation time, consistent with solid-like characteristics[58]. In contrast, the ratio of $G'/G''$ of Layer-2 and Layer-3 decrease by 26% and 38%, respectively. The attractive interactions are relatively weaker in Layer-2 and Layer-3 compared to Layer-1, and thus entropic repulsion becomes significant in these outer layers[20,32]. Since the Layer-2 and Layer-3, located approximately 2-3 nm from the substrate, are not in direct contact with the surface, allowing polymer chains to retain greater conformational freedom. Over time, these segments experience increasing excluded volume interactions and tend to maximize their conformational entropy. This gives rise to effectively repulsive forces between chains, commonly referred to as entropic repulsion[32]. Consequently, these upper layers exhibit fluid-like rheological behavior, characterized by shorter relaxation times and reduced storage modulus. As a result, polymers at Layer-2 and Layer-3 gain degrees of freedom, undergo bulk entanglement, and behave as viscous fluids with increased viscosity over time. As physical aging progresses, the interfacial structure becomes stratified, with entanglement-induced solidification dominating near the surface, while entropic repulsion governs the rheology of the outer layers.

These findings not only provide fundamental insight into the time-dependent solidification behavior of polymeric fluids under nanoconfinement but also suggest potential applications in relevant technological fields. In particular, the emergence of metastable interfacial solid-like layers may have implications for nanoscale lubrication control where aging-dependent mechanical responses influence sliding performance[59]. Similarly, the time-dependent viscoelastic properties may inform the design of adaptive adhesion surfaces[60].

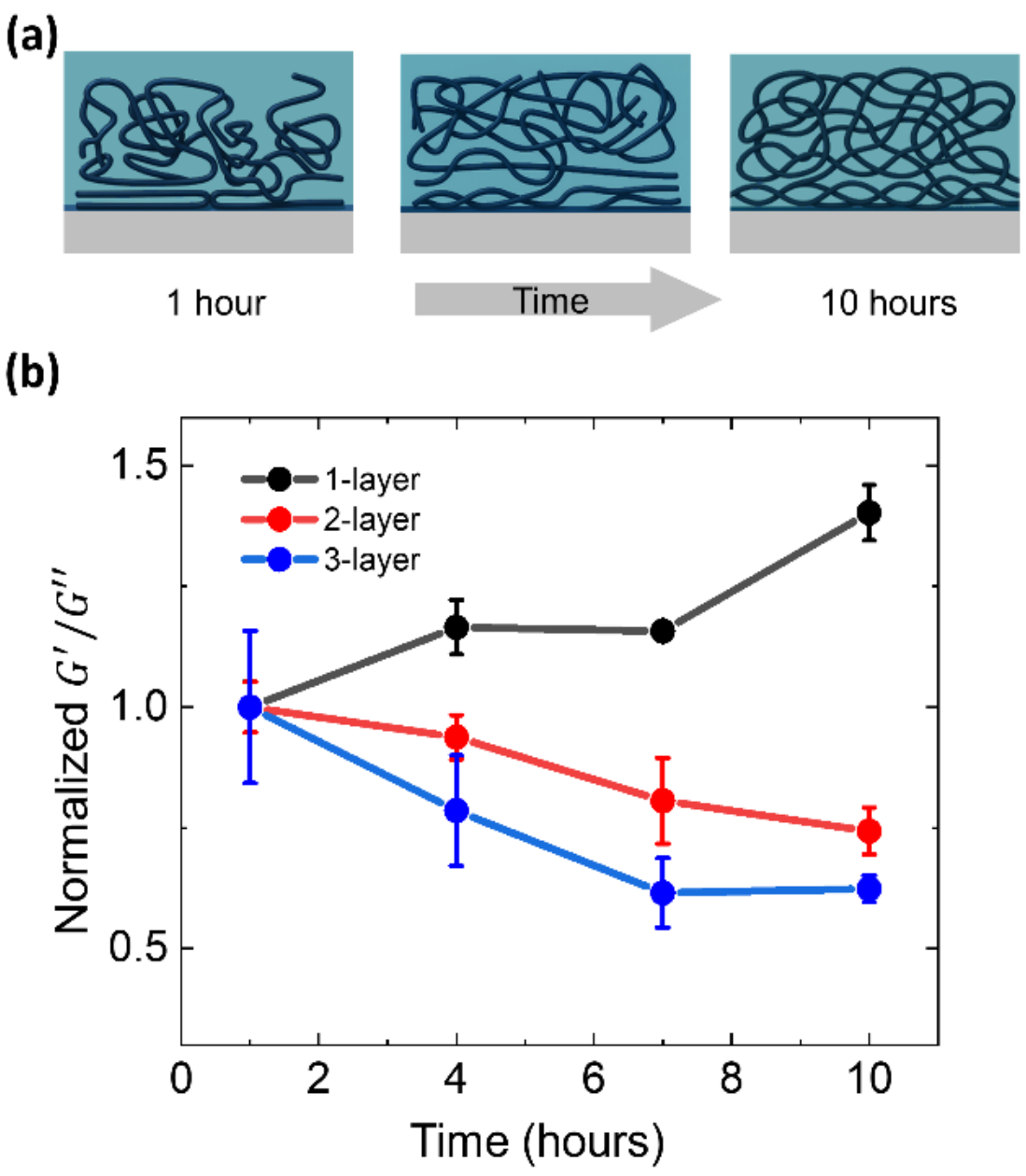

**Fig. 3.** (a) Schematic illustration of physical aging of polymer structure on mica substrate. Initially formed layered structures evolve over time. After 10 hours, polymer adsorption at the solid interface enhances interfacial entanglement, leading to the solidification of polymers at the bottommost layer. (b) Normalized ratio $G'/G''$ of 18,000 g/mol (500 cSt) PDMS-based silicone oil on mica. Each layer is normalized to an initial value at 1 hour. Over 10 hours of aging, the $G'/G''$ of Layer-1 increases by approximately 40%, while those of Layer-2 and Layer-3 decrease by 26% and 38%, respectively. This suggests increased solidity in Layer-1 and reduced solidity in the outer layers.

## IV. Slip boundary changes due to interfacial entanglement

The interfacial entanglement-induced solidification alters the slip boundary over time, as shown in Fig. 4(a). To characterize the nanoscale variation of slip boundary, we measure the lateral damping force and analyze the slip length of the flow[61,62]. Typically, the fluid velocity at a solid surface is assumed to be zero, but if the fluid moves with a finite velocity at the interface, it is characterized by a positive-valued slip length[61,63]. In contrast, when the fluid forms an immobilized interfacial film, the fluid dynamics at the interface can be consistently described by a negative-valued slip length[62,64]. We observe that the flow boundaries progressively extend upward away from the substrate, which is consistent with the solidification of Layer-1 with about 1 nm thickness. This further supports the presence of solid-like interfacial polymers at Layer-1.

The physical aging process induces a progressive shift in the flow boundary, as measured through our QTF-AFM-based force spectroscopy (Fig. 4(b)). This shift is attributed to interfacial entanglement occurring in the bottommost layer. We can determine the slip length for the confined fluids from the fit based on Maxwell model (Eq.(8))[61,63]. At larger tip–sample separations, the lateral damping force decreases as the shear interaction gradually vanishes to zero. We fit the experimentally measured data to the model force in the range from 30 nm down to 1 nm. Below 1 nm, molecular dimensions become significant, and the classical continuum description based on the Navier-Stokes equations breaks down[5]. The dashed vertical lines, located below 1 nm, indicate the slip boundary; the length $b$ is extracted from the fitted damping force profiles. Unlike Newtonian fluids such as water, which exhibit often slip and fluid-like behavior even at 1 nm from solid surface depending on surface wettability[61], the polymeric fluid in this study undergoes solidification near the interface during aging. These solidified polymers act as a thin film with monomer-scale thickness on the solid surface, leading to increasingly negative slip lengths over time[62,64].

The slip length is initially measured to be slightly negative at $-0.25 \pm 0.15$ nm (1-hour). As aging proceeds, the flow boundary expands approximately threefold, reaching $-0.60 \pm 0.03$ nm. The negative value of slip length indicates that the effective no-slip boundary plane is shifted into the fluid, away from the solid surface[62,64,65]. This suggests that the fluid adheres strongly to the solid surface and solidifies through interfacial entanglement in the first polymer layer[66]. As the solid-like (Layer-1) and fluid-like (Layer-2 and -3) behaviors become more pronounced (Fig. 3(b)), the magnitude of the slip length reaches the monomer diameter ~0.8 nm. Notably, this trend is consistent with the normalized $G'/G''$ of Layer-1 (Fig. 3(b)).

While this study focuses on PDMS due to its well-characterized viscoelasticity and widespread use in rheological investigations, the underlying mechanism of interfacial entanglement-induced solidification may extend to other entangled polymer systems. For example, in Polystyrene, Poly(2-vinylpryridine), Poly(methyl methacrylate), irreversible adsorption onto solid walls leading to entanglement and a significant increase in segmental relaxation time, as observed *via* dielectric spectroscopy[58]. Similarly, in hydrogel system, pronounced stiffening and shear modulus enhancement under confinement have been attributed to polymer entanglements[54]. Although our work is the first to directly quantify nanoscale rheological behavior of polymers at solid interfaces, the qualitative agreement with these earlier structural and optical studies suggests that this entanglement-induced solidification may be broadly relevant to other polymer systems. This highlights a promising avenue for extending nanorheological approaches to diverse material platforms. In particular, the slip length becoming increasingly negative over time implies a significant enhancement in interfacial flow resistance. This phenomenon is especially relevant to micro/nanofluidic systems, where even nanometer-scale modifications in boundary conditions can dramatically alter transport properties. After 10 hours of aging, the bottommost interfacial layer behaves as a tightly adsorbed, monolayer-like film with a fivefold increase in storage modulus ($0.577 \pm 0.179$ MPa), effectively narrowing the flow channel and impeding shear flow near the solid boundary. Such development of a solid-like interfacial layer leads to a negative slip lengths, as shown in Fig. 4(c), and reflects in other nanoconfined systems such as shale reservoirs, where deviations from classical Darcy flow arise due to similar effects under confinement[67]. Thus, the interfacial entanglement-induced solidification provides molecular level mechanism for flow resistance in confined polymeric systems.

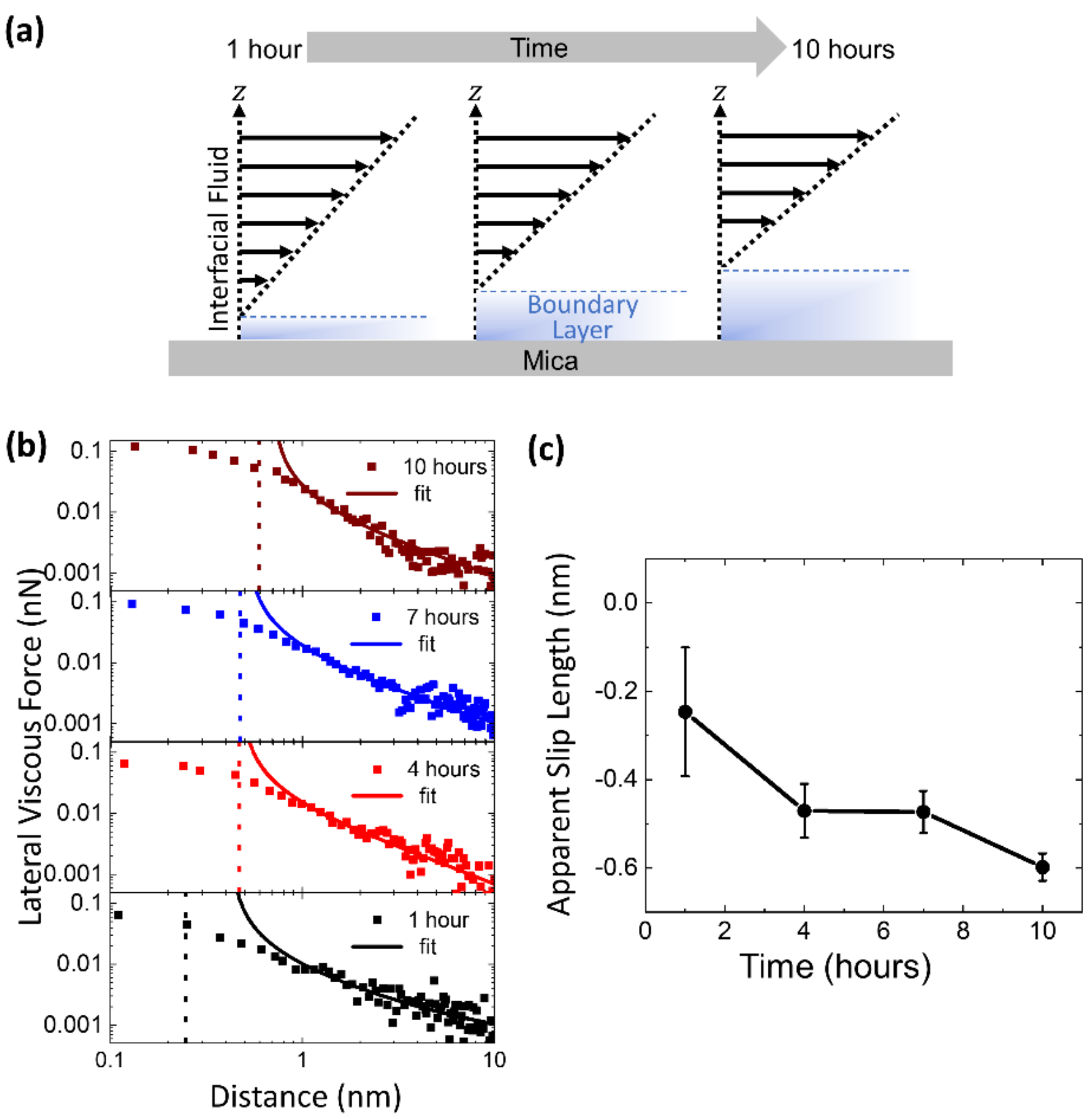

**Fig. 4.** (a) Schematic illustration of flow profiles for interfacial fluids within 5 nm of the solid interface. At 1 hour, a nearly no-slip boundary condition is observed. After 10 hours, adsorption-induced interfacial entanglement leads to an upward shift of the flow boundary, forming a strongly bound solid-like polymer film on the solid surface. (b) Temporal evolution of the flow boundary, as indicated by the vertical dashed lines. Lateral damping forces are measured as a function of distance and fitted using the Maxwell model, plotted on a log-log scale. The fitting range spans from 30 nm to 1 nm. (c) Evolution of the slip length over time, with the slip lengths obtained by averaging eight force data at the respective layer peak positions within ±0.05 nm. The slip length becomes increasingly negative, reflecting the upward shift of the flow boundary.

## V. Conclusions

We have investigated the time-dependent structure and rheological properties of interfacial polymers using a home-built quartz tuning fork–atomic force microscope-based rheometer. Initially, fluid molecules form pronounced layered structures. However, the structure evolves over 10 hours, while altering properties; the storage modulus of bottommost layer (Layer-1) increases fivefold compared to the 1-hour value—reaching 16 times the storage modulus of the bulk. This indicates a transition toward solid-like behavior. In contrast, Layer-2 and Layer-3 persist with relatively stable storage moduli and their loss moduli substantially increase, indicating a transition toward fluid-like behavior.

We attribute the observed solidification in Layer-1 to adsorption-driven polymer entanglement onto the substrate, mediated by attractive interactions between PDMS chains and the mica surface. As aging proceeds, polymer chains near the interface become partially immobilized through hydrogen bonding, forming a tightly bound interfacial network. The earliest arriving chains tend to flatten to maximize contact area, while later-arriving chains dynamically entangle by loops and tails. This irreversible adsorption leads to an interfacial entanglement that exhibits solid-like mechanical response. Notably, the solidification process is not by crystal nucleation. In contrast, Layer-2 and Layer-3, located 2–3 nm above the substrate, do not directly interact with the solid surface. Over time, these polymers experience increased excluded volume effects and tend to maximize their conformational entropy, giving rise to entropic repulsion. This promotes fluid-like behavior, reflected in the increasing loss modulus and decreasing relaxation time.

After 10 hours of aging, the bottommost layer evolves into a densely adsorbed monolayer-like film with markedly enhanced rigidity. This structural transition effectively narrows the flow channel and increases resistance to interfacial shear. The corresponding rheological response, characterized by an increased storage modulus and a prolonged relaxation time, is consistent with the emergence of negative slip lengths. These results support a unified view in which interfacial entanglement, advancement of solid-liquid interface, and time-dependent solidification govern the behavior of polymers at solid interfaces.

**References**

[1] H.K. Christenson, R.G. Horn, and J.N. Israelachvili, "Measurement of forces due to structure in hydrocarbon liquids," Journal of Colloid and Interface Science **88**(1), 79–88 (1982).
[2] M. Heuberger, M. Zäch, and N.D. Spencer, "Density Fluctuations Under Confinement: When Is a Fluid Not a Fluid?," Science **292**(5518), 905–908 (2001).
[3] R.G. Horn, and J.N. Israelachvili, "Direct measurement of structural forces between two surfaces in a nonpolar liquid," The Journal of Chemical Physics **75**(3), 1400–1411 (1981).
[4] J.N. Israelachvili, *Intermolecular and Surface Forces* (Academic Press, 2011).
[5] N. Kavokine, R.R. Netz, and L. Bocquet, "Fluids at the Nanoscale: From Continuum to Subcontinuum Transport," Annu. Rev. Fluid Mech. **53**(1), 377–410 (2021).
[6] A. Maali, T. Cohen-Bouhacina, G. Couturier, and J.-P. Aimé, "Oscillatory Dissipation of a Simple Confined Liquid," Phys. Rev. Lett. **96**(8), 086105 (2006).
[7] S.H. Khan, G. Matei, S. Patil, and P.M. Hoffmann, "Dynamic Solidification in Nanoconfined Water Films," Phys. Rev. Lett. **105**(10), 106101 (2010).
[8] J. Gao, W.D. Luedtke, and U. Landman, "Layering Transitions and Dynamics of Confined Liquid Films," Phys. Rev. Lett. **79**(4), 705–708 (1997).
[9] P.G. Debenedetti, *Metastable Liquids: Concepts and Principles* (Princeton University Press, 2020).
[10] A.L. Demirel, and S. Granick, "Glasslike Transition of a Confined Simple Fluid," PHYSICAL REVIEW LETTERS **77**(11), (1996).
[11] A.K. Doerr, M. Tolan, J.-P. Schlomka, and W. Press, "Evidence for density anomalies of liquids at the solid/liquid interface," Europhys. Lett. **52**(3), 330–336 (2000).
[12] Q. Yang, P.Z. Sun, L. Fumagalli, Y.V. Stebunov, S.J. Haigh, Z.W. Zhou, I.V. Grigorieva, F.C. Wang, and A.K. Geim, "Capillary condensation under atomic-scale confinement," Nature **588**(7837), 250–253 (2020).
[13] S. Benaglia, M.R. Uhlig, J. Hernández-Muñoz, E. Chacón, P. Tarazona, and R. Garcia, "Tip Charge Dependence of Three-Dimensional AFM Mapping of Concentrated Ionic Solutions," Phys. Rev. Lett. **127**(19), 196101 (2021).
[14] D.F. Kienle, and T.L. Kuhl, "Density and Phase State of a Confined Nonpolar Fluid," Phys. Rev. Lett. **117**(3), 036101 (2016).
[15] L. Grill, and S. Hecht, "Covalent on-surface polymerization," Nat. Chem. **12**(2), 115–130 (2020).
[16] T.P. Russell, and Y. Chai, "*50th Anniversary Perspective* : Putting the Squeeze on Polymers: A Perspective on Polymer Thin Films and Interfaces," Macromolecules **50**(12), 4597–4609 (2017).
[17] G. Evmenenko, H. Mo, S. Kewalramani, and P. Dutta, "Conformational rearrangements in interfacial region of polydimethylsiloxane melt films," Polymer **47**(3), 878–882 (2006).
[18] G. Evmenenko, S.W. Dugan, J. Kmetko, and P. Dutta, "Molecular Ordering in Thin Liquid Films of Polydimethylsiloxanes," Langmuir **17**(13), 4021–4024 (2001).
[19] R.G. Horn, and J.N. Israelachvili, "Molecular organization and viscosity of a thin film of molten polymer between two surfaces as probed by force measurements," Macromolecules **21**(9), 2836–2841 (1988).
[20] G. Sun, M. Kappl, T. Pakula, K. Kremer, and H.-J. Butt, "Equilibrium Interaction of Solid Surfaces across a Polymer Melt," Langmuir **20**(19), 8030–8034 (2004).
[21] T. Fukuma, and R. Garcia, "Atomic- and Molecular-Resolution Mapping of Solid–Liquid Interfaces by 3D Atomic Force Microscopy," ACS Nano **12**(12), 11785–11797 (2018).
[22] K. Amano, K. Suzuki, T. Fukuma, O. Takahashi, and H. Onishi, "The relationship between local liquid density and force applied on a tip of atomic force microscope: A theoretical analysis for simple liquids," The Journal of Chemical Physics **139**(22), 224710 (2013).
[23] H.P. Kavehpour, and G.H. McKinley, "Tribo-Rheometry: From Gap-Dependent Rheology to Tribology," Tribology Letters **17**(2), 327–335 (2004).
[24] B. Bhushan, J.N. Israelachvili, and U. Landman, "Nanotribology: friction, wear and lubrication at the atomic scale," Nature **374**(6523), 607–616 (1995).
[25] H.-J. Butt, J. Wang, R. Stark, M. Kappl, B.A. Wolf, J. Eckelt, and A. Knopf, "Forces Between Solid Surfaces Across Polymer Melts as Revealed by Atomic Force Microscopy," Soft Materials **5**(2–3), 49–60 (2007).

[26] V.S.J. Craig, "An historical review of surface force measurement techniques," Colloids and Surfaces A: Physicochemical and Engineering Aspects **129–130**, 75–93 (1997).
[27] O. Gbadeyan, S. Adali, G. Bright, B. Sithole, and P. Lekha, "Mechanical, microstructure, and dynamic mechanical analysis of nano-shell and plant fiber hybrid biocomposite," Journal of Composite Materials **55**(24), 3345–3358 (2021).
[28] M. Lee, B. Sung, N. Hashemi, and W. Jhe, "Study of a nanoscale water cluster by atomic force microscopy," Faraday Discuss. **141**, 415–421 (2009).
[29] Y. Seo, and W. Jhe, "Atomic force microscopy and spectroscopy," Rep. Prog. Phys. **71**(1), 016101 (2007).
[30] H.R. Brown, and T.P. Russell, "Entanglements at Polymer Surfaces and Interfaces," Macromolecules **29**(2), 798–800 (1996).
[31] A. Pawlak, and J. Krajenta, "Entanglements of Macromolecules and Their Influence on Rheological and Mechanical Properties of Polymers," Molecules **29**(14), 3410 (2024).
[32] Y. Nishiwaki, Y. Yamada, T. Utsunomiya, H. Sugimura, and T. Ichii, "Interfacial Structures and Mechanical Response of Highly Viscous Polymer Melt on Solid Surfaces Investigated by Atomic Force Microscopy," J. Phys. Chem. C **128**(28), 11966–11976 (2024).
[33] F.J. Giessibl, "High-speed force sensor for force microscopy and profilometry utilizing a quartz tuning fork," Applied Physics Letters **73**(26), 3956–3958 (1998).
[34] J. Shim, C. Kim, M. Lee, S. An, and W. Jhe, "Multiscale rheology from bulk to nano using a quartz tuning fork-atomic force microscope," Review of Scientific Instruments **95**(10), 105104 (2024).
[35] S.H.Jr. Donaldson, S. Das, M.A. Gebbie, M. Rapp, L.C. Jones, Y. Roiter, P.H. Koenig, Y. Gizaw, and J.N. Israelachvili, "Asymmetric Electrostatic and Hydrophobic–Hydrophilic Interaction Forces between Mica Surfaces and Silicone Polymer Thin Films," ACS Nano **7**(11), 10094–10104 (2013).
[36] S. Jiang, S.C. Bae, and S. Granick, "PDMS Melts on Mica Studied by Confocal Raman Scattering," Langmuir **24**(4), 1489–1494 (2008).
[37] G. Sun, and H.-J. Butt, "Adhesion between Solid Surfaces in Polymer Melts: Bridging of Single Chains," Macromolecules **37**(16), 6086–6089 (2004).
[38] C. Kim, M.C. Gurau, P.S. Cremer, and H. Yu, "Chain Conformation of Poly(dimethyl siloxane) at the Air/Water Interface by Sum Frequency Generation," Langmuir **24**(18), 10155–10160 (2008).
[39] H.(王海鹏) Wang, D.(刘丁楠) Liu, C.(郑晨辉) Zheng, J.(赵炯飞) Zhao, J.(常健) Chang, L.(胡亮) Hu, H.(廖晖) Liao, D.(耿德路) Geng, W.(解文军) Xie, and B.(魏炳波) Wei, "Spiral eutectic growth dynamics facilitated by space Marangoni convection and liquid surface wave," Physics of Fluids **36**(4), 047137 (2024).
[40] S. Kim, D. Kim, J. Kim, S. An, and W. Jhe, "Direct Evidence for Curvature-Dependent Surface Tension in Capillary Condensation: Kelvin Equation at Molecular Scale," Phys. Rev. X **8**(4), 041046 (2018).
[41] R.G. Larson, "The Structure and Rheology of Complex Fluids," (n.d.).
[42] B. Kim, J. Jahng, R.M. Khan, S. Park, and E.O. Potma, "Eigenmodes of a quartz tuning fork and their application to photoinduced force microscopy," Phys. Rev. B **95**(7), 075440 (2017).
[43] E.A. López-Guerra, and S.D. Solares, "Modeling viscoelasticity through spring–dashpot models in intermittent-contact atomic force microscopy," Beilstein J. Nanotechnol. **5**(1), 2149–2163 (2014).
[44] J. Kim, D. Won, B. Sung, and W. Jhe, "Observation of Universal Solidification in the Elongated Water Nanomeniscus," J. Phys. Chem. Lett. **5**(4), 737–742 (2014).
[45] S. Jeffery, P.M. Hoffmann, J.B. Pethica, C. Ramanujan, H.Ö. Özer, and A. Oral, "Direct measurement of molecular stiffness and damping in confined water layers," Phys. Rev. B **70**(5), 054114 (2004).
[46] M. Lee, J. Hwang, B. Kim, S. An, and W. Jhe, "Fluid-induced resonances in vibrational and Brownian dynamics of a shear oscillator," Current Applied Physics **16**(11), 1459–1463 (2016).
[47] A.J. Barlow, G. Harrison, J. Lamb, and J.M. Robertson, "Viscoelastic relaxation of polydimethylsiloxane liquids," Proceedings of the Royal Society of London. Series A. Mathematical and Physical Sciences **282**(1389), 228–251 (1964).
[48] B. Kim, S. Kwon, H. Mun, S. An, and W. Jhe, "Energy dissipation of nanoconfined hydration layer: Long-range hydration on the hydrophilic solid surface," Sci Rep **4**(1), 6499 (2014).
[49] B. Kim, S. Kwon, M. Lee, Qh. Kim, S. An, and W. Jhe, "Probing nonlinear rheology layer-by-layer in interfacial hydration water," Proc. Natl. Acad. Sci. U.S.A. **112**(51), 15619–15623 (2015).
[50] C. Nannette, J. Baudry, A. Chen, Y. Song, A. Shglabow, N. Bremond, D. Démoulin, J. Walters, D.A. Weitz, and J. Bibette, "Thin adhesive oil films lead to anomalously stable mixtures of water in oil," Science **384**(6692), 209–213 (2024).
[51] T. Bardelli, C. Marano, and F. Briatico Vangosa, "Polydimethylsiloxane crosslinking kinetics: A systematic

study on Sylgard184 comparing rheological and thermal approaches," J of Applied Polymer Sci **138**(39), 51013 (2021).
[52] H.I. Jeong, H.S. Jung, M. Dubajic, G. Kim, W.H. Jeong, H. Song, Y. Lee, S. Biswas, H. Kim, B.R. Lee, J.W. Yoon, S.D. Stranks, S.M. Jeong, J. Lee, and H. Choi, "Super elastic and negative triboelectric polymer matrix for high performance mechanoluminescent platforms," Nat Commun **16**(1), 854 (2025).
[53] M. Rubinstein, and R.H. Colby, *Polymer Physics* (Oxford University Press, 2003).
[54] C. Liang, V. Dudko, O. Khoruzhenko, X. Hong, Z.-P. Lv, I. Tunn, M. Umer, J.V.I. Timonen, M.B. Linder, J. Breu, O. Ikkala, and H. Zhang, "Stiff and self-healing hydrogels by polymer entanglements in co-planar nanoconfinement," Nat. Mater. **24**(4), 599–606 (2025).
[55] Z. Wang, M. Schaller, A. Petzold, K. Saalwächter, and T. Thurn-Albrecht, "How entanglements determine the morphology of semicrystalline polymers," Proc. Natl. Acad. Sci. U.S.A. **120**(27), e2217363120 (2023).
[56] R.S. Lakes, *Viscoelastic Solids (1998)*, First edition (CRC Press, Boca Raton, FL, 2017).
[57] S. Patil, G. Matei, A. Oral, and P.M. Hoffmann, "Solid or Liquid? Solidification of a Nanoconfined Liquid under Nonequilibrium Conditions," Langmuir **22**(15), 6485–6488 (2006).
[58] S. Napolitano, and M. Wübbenhorst, "The lifetime of the deviations from bulk behaviour in polymers confined at the nanoscale," Nat Commun **2**(1), 260 (2011).
[59] S. Liu, D. Guo, and G. Xie, "Nanoscale lubricating film formation by linear polymer in aqueous solution," Journal of Applied Physics **112**(10), 104309 (2012).
[60] C. Linghu, R. Wu, Y. Chen, Y. Huang, Y.-J. Seo, H. Li, G. Wang, H. Gao, and K.J. Hsia, "Long-term adhesion durability revealed through a rheological paradigm," Sci. Adv. **11**(11), eadt3957 (2025).
[61] D. Ortiz-Young, H.-C. Chiu, S. Kim, K. Voïtchovsky, and E. Riedo, "The interplay between apparent viscosity and wettability in nanoconfined water," Nat Commun **4**(1), 2482 (2013).
[62] A.A. Shuvo, L.E. Paniagua-Guerra, J. Choi, S.H. Kim, and B. Ramos-Alvarado, "Hydrodynamic slip in nanoconfined flows: a review of experimental, computational, and theoretical progress," Nanoscale **17**(2), 635–660 (2025).
[63] L.D. Landau, and E.M. Lifshitz, *Fluid Mechanics: Volume 6* (Elsevier, 1987).
[64] H.-J. Butt, and M. Kappl, *Surface and Interfacial Forces* (John Wiley & Sons, 2018).
[65] Y. Luo, A.-P. Pang, and X. Lu, "Liquid-Solid Interfaces under Dynamic Shear Flow: Recent Insights into the Interfacial Slip," Langmuir **38**(15), 4473–4482 (2022).
[66] O. Bäumchen, R. Fetzer, and K. Jacobs, "Reduced Interfacial Entanglement Density Affects the Boundary Conditions of Polymer Flow," Phys. Rev. Lett. **103**(24), 247801 (2009).
[67] R. Wang, J. Chai, B. Luo, X. Liu, J. Zhang, M. Wu, M. Wei, and Z. Ma, "A review on slip boundary conditions at the nanoscale: recent development and applications," Beilstein J. Nanotechnol. **12**, 1237–1251 (2021).

## Data availability

The data sets generated and/or analyzed during the current study are available from the corresponding authors on reasonable request.

## Supplementary Material

See the supplementary material for the layering interval determinations, Maxwell model based lateral damping force and mechanical relaxation analysis are included.

## Acknowledgements

The authors acknowledge Prof. D. Weitz and Prof. G. McKinley for helpful discussions. The authors also thank Dr. Chungman Kim for valuable discussions during the early stages of this work. The SEM images of the tip were obtained at the National Instrumentation Center for Environmental Management (NICEM), Seoul National University. This work was supported by the National Research Foundation of Korea (NRF) grant funded by the Korean government (MSIP) (No. 2016R1A3B1908660) and the Commercialization Promotion Agency for R&D Outcomes (COMPA) funded by the Ministry of Science and ICT(MSIT) (1711198537).

## Author Declarations

## Conflict of Interest

The authors have no conflicts to disclose.

## Author Contributions

Jaewon Shim: Data curation, Writing – original draft. Manhee Lee: Writing – original draft, Supervision (equal). Wonho Jhe: Conceptualization (lead); Funding acquisition (equal).